\documentclass[showpacs,preprintnumbers,amsmath,amssymb]{revtex4}
\usepackage{graphicx}
\usepackage{amsfonts}
\usepackage{amsmath}
\usepackage{color}
\usepackage{wrapfig,floatflt}
\newcommand{\beq}{\begin{equation}}
\newcommand{\eneq}{\end{equation}}
\newcommand{\bea}{\begin{eqnarray}}
\newcommand{\enea}{\end{eqnarray}}
\newcommand{\beqs}{\begin{subequations}}
\newcommand{\eeqs}{\end{subequations}}

\begin{document}

\title{Entanglement in a spin system with inverse square statistical interaction}

\author{D. Giuliano , A. Sindona, G. Falcone, F. Plastina}

\affiliation{Dipartimento di Fisica, Universit\`a della Calabria and
             I.N.F.N., Gruppo collegato di Cosenza, Arcavacata di Rende
             I-87036, Cosenza, Italy}

\author{L. Amico}

\affiliation{MATIS-INFM \& Dipartimento di Metodologie Fisiche e Chimiche (DMFCI),
Universit\`a di Catania, viale A. Doria 6, 95125 Catania, Italy}


\date{\today}
\begin{abstract}
We investigate the entanglement content of the ground state of a
system characterized by effective elementary  degrees of freedom
with fractional statistics. To this end, we explicitly construct
the ground state for a chain of $N$ spins with inverse square
interaction~(the Haldane-Shastry model) in the presence of an
external uniform magnetic field. For such a system at zero
temperature, we evaluate the entanglement in the ground state both
at finite size and in the thermodynamic limit. We relate  the
behavior of the quantum correlations with the spinon condensation
phenomenon occurring at the saturation field.
\end{abstract}

\maketitle

\section{Introduction}
Quantum states of interacting many body systems are inherently
entangled. This simple statement has motivated a wide physics
community to study, in the light of quantum information theory,
models typically explored in quantum statistical
mechanics~\cite{amicorep}. Besides the motivation of the crucial
role played by the entanglement as a resource  for quantum
computation protocols, the promise is to acquire a deeper
understanding of the physical systems themselves, with possible
spin off on open problems in condensed matter. There are
indications that, while local interactions are possibly the most
common mechanisms leading to non local correlations, also
statistics  can produce entanglement. Indeed, because of the
symmetrization constraints, most generically  identical particles 
are not in a product state. Even though the resulting many body
wave function cannot lead to  genuine entangled state because of
the lack of the tensor product structure in Fock
spaces~\cite{Ghirardi}, Vedral~\cite{Vlatko-fermions} demonstrated
that a free fermion gas enjoys spin entanglement on distances of
the order of the inverse Fermi momentum; on the contrary, no such
kind of entanglement has been found for the polarizations of identical 
bosons. An alternative route to capture non local correlations in
the states of identical particles was pursued by Zanardi and
coworkers through the so called \emph{mode-entanglement}, directly
defined in the Fock space~\cite{Zanardi}. As a basic support to
this point of view, one can argue that separable states in the
dual space correspond to configurationally entangled states.
Furthermore, in favor of such  `statistics entanglement', it should
be noticed that several protocols have been suggested to exploit
it for computational purposes~\cite{Cavalcanti}.

In this paper, we deal with statistics in a one dimensional
interacting many particle system. In one dimension, statistics is
peculiar as particles need to scatter each other in order to
exchange their position. Accordingly,  one dimensional statistics
can be formulated as a boundary condition for the many-body wave
function in the configurational space~\cite{Les-Houches1999}. In
this context, particularly relevant are the Calogero-Sutherland
models~(CSM)s, i.e., quantum mechanical models in one dimension
where the interaction among particles is inversely proportional to
the square of their distance. Indeed,  such models can be regarded
as describing a gas of identical free particles with fractional
statistics~\cite{Haldane-exclusion,Calogero-statistics}, which has
made them relevant in the understanding of the fractional quantum
Hall effect.

The main motivation of our work is to investigate the kind of
entanglement emerging from fractional statistics.
We note that, because of the lack of any tensor product structure, the 
entanglement {\it between} particles with fractional statistics has not been fully quantified yet\cite{partitions}.
We concentrate on the spin entanglement
emerging from quasi-particles with fractional
statistics. To rely on some tensor product
structures of the underlying Hilbert space hosting the quantum
system, we consider the Haldane-Shastry~\cite{haldaneshastry}
model~(HSM): an anti-ferromagnetic spin-$1/2$ chain  with inverse
square exchange interaction among the spins.

Also the HSM~(in the dual space) can be interpreted as an
ideal  gas with fractional statistics~\cite{Haldane-ideal}. The
connection with the CSM can be put on firm grounds following the
scheme developed by Polychronakos; namely, first introducing some
internal degrees of freedom into the CS model and then employing
the so called `freezing trick', to  fix the particle positions~\cite{toadd1}.
As can be understood by a direct inspection of the
eigenstates of the two models, it is worth mentioning that the
correspondence between the CSM and the HSM holds only for a
well-defined value of the parameter characterizing the statistics
of the excitations in the CSM,  identifying a `semion' gas (i.e.,
a gas of particles with statistics half than the Fermi statistics).
One of the most spectacular evidences that the HSM is indeed an ideal
gas of semions is provided by the response of the system to a
local perturbation: it can be proved that a single spin flip
causes the creation of a pair of collective excitations~(spinons).
Higher `angular momenta' excitations are generated by two or more
spin flips~\cite{haldanemagnetic}.

In the present paper, we study the entanglement in the ground
state of the HSM in an external magnetic field. In fact, the
applied magnetic field causes  a certain number of spin flips to
occur in the ground state: this allows us to study quantum
correlations in the presence of spinon condensation. We shall
demonstrate that the divergence of the spinon 'Fermi' wavelength
occurring when all the spinons in the system actually condense
(saturation field), a characteristic trait of the semionic
statistics,  causes a divergence of the entanglement length. We
also note  that in our work we study the interplay between the
range of local interaction and the range of the entanglement. We
eventually find that a specific pattern of entanglement
characterizes the system with a finite range interaction.

The paper is organized as follows: in the following section,
we introduce the model and explicitly construct its ground
state as function of the magnetic field; in Sect.~\ref{entangle},
we employ the single- and two-spin ground state averages,
computed in Appendix~\ref{spinprop} to evaluate various entanglement
measures. Finally, Sect.~\ref{close} contains some concluding
remarks.

\section{The model}

The Haldane-Shastry~(HS) Hamiltonian~\cite{haldaneshastry}
describes a one-dimensional spin-$1/2$ chain wrapped around a
circle of unit radius,  with an anti-ferromagnetic interaction
whose strength is inversely proportional to the square of the
chord between the corresponding sites. When the model is placed in
an external uniform magnetic field $h$, its Hamiltonian is given
by
\beq H=\frac{ 2
\pi^2 J}{N^2} \: \sum_{\alpha \neq \beta = 0}^{N-1} \: \frac{ {\bf
S}_\alpha \cdot {\bf S}_\beta }{ | z_\alpha - z_\beta |^2 } +  h
\sum_{ \alpha = 0}^{N-1} S_\alpha^z \:\:\:\: .
\label{model1}
\eneq
\noindent In Eq.~(\ref{model1}), ${\bf S}_\alpha$ is the spin
operator and $z_{\alpha}$ the spin coordinate of the site $\alpha = 0 , 1 , \ldots , N-1$, with
$N$ the total number of spins.
Periodically boundary conditions $z_{\alpha} = z_{\alpha+N}$, and
 ${\bf S}_\alpha = {\bf S}_{\alpha + N}$, allow to parameterize the
positions on the circle by the $N^{\rm th}$ roots of the unity $z_\alpha =  e^{
2 \pi i \alpha / N}$.
For $h=0$,
the ground state of the system is a non-degenerate spin liquid
with $0$ total spin (a spin singlet)~\cite{haldaneshastry}.
At zero field, flipping
a spin in the ground state corresponds to creating a~(fully polarized) pair of spin-$1/2$ collective excitations~\cite{faddeev},
dubbed ``spinons'' by Haldane~\cite{Haldane-exclusion}.
The spinons  keep their integrity  when scattered off each other,
so that they can be thought of as true quantum particles, despite
their collective nature.
Any time one more spin is flipped, an
additional spinon pair is created in the state of the system;
flipping $K/2$ spins is, thus, equivalent to creating a fully
polarized $K$-spinon excited state.
The Zeeman energy in
Eq.~(\ref{model1}) lowers the energy of the fully polarized states,
since  $h$ works like a chemical potential for the spinons.
Thus,
upon increasing the magnetic field the (even)~number of spinons $K$ increases
and one of the fully polarized $K$-spinon
states becomes the ground state of the system.
This state can be written as
\beq
| \Phi_{\rm gnd} , K  \rangle = \sum_{ z_1 , \ldots , z_M }
\: \Phi_K ( z_1 , \ldots , z_M ) \: | z_1 , \ldots , z_M \rangle
\:\:\:\: ,
\label{model2}
\eneq
\noindent where
$z_1 , \ldots , z_M$ denote the positions of the $\uparrow$-spins
in the state, the remaining ones pointing downwards.
In Eq.~(\ref{model2}),
the number of $\uparrow$-spins is given by
$M = (N - K )/2$, where $K$ is the number of condensed $\downarrow$-spinons
in the ground state;
the
state $ | z_1, \ldots , z_M \rangle $ is understood
to be equal to
$ | z_1, \ldots , z_M \rangle = S_1^+ \ldots S_M^+ \: | \downarrow
\rangle^{\otimes N}$
and the amplitude $\Phi_K ( z_1 , \ldots , z_M
)$ has the form
\beq
\Phi_K ( z_1 , \ldots , z_M ) = \left[ \frac{N^M (2M)!}{ 2^M}
\right]^{ - \frac{1}{2}} \: \prod_{ t = 1}^M z_t^{ 1 + K /2} \:
\prod_{ i < j =1}^M ( z_i - z_j )^2 \:\:\:\: .
\label{model3}
\eneq
\noindent The number of condensed spinons is determined by
the value of the external field.
In order to find out how $K$
increases with $h$, one has first to calculate
the energy $ E_{\rm gnd} [ N , K , h]$
of
the state $ | \Phi_{\rm gnd} , K \rangle $, as a function of $K$
and $h$ at fixed $N$, and then to
determine the value of $K$ that minimizes $ E_{\rm gnd} [ N , K ,
h]$, at fixed $h$.
$ E_{\rm gnd} [ N , K , h]$  can be obtained, for
example, by a straightforward generalization of the approach
developed in Ref.~\cite{bgl}:
since
$\Phi_K ( z_1 , \ldots , z_M)$ is a homogeneous polynomial
of $z_1 , \ldots , z_M$, the summations over spin indices, contained in the expectation value
$\langle  \Phi_{\rm gnd} , K  | H |  \Phi_{\rm gnd} , K \rangle$,
are
replaced by
derivative operators that are understood to act
onto the analytic extension of $\Phi_{K} ( z_1 , \ldots , z_{M})$,
in which the spin coordinates are allowed to take any value on the unit
circle. After computing the derivatives, the variables $z_1 ,
\ldots , z_M$ are constrained back to the lattice sites.
The ground
state energy is then the minimum energy of a $K$-spinon condensate
at finite $h$; it can be expressed as
\beq
E_{\rm gnd} [ N , K , h] =
E [ q_1 ( M ) , \ldots ,q_{\frac{K}{2}} ( M ), q_{ \frac{K}{2} + 1} ( 0 ) , \ldots q_K ( 0
)] - h \frac{K}{2} \;,
\label{model7}
\eneq
in which
\beq
E [\ldots, q_\ell (m_{\ell}), \ldots] = - \frac{\pi^2 J ( N^2 + 5 )}{ 24 N} + \frac{1}{2}
\sum_{\ell = 1}^M \left [ J \left( \frac{\pi}{2} \right)^2 - J
q_\ell(m_{\ell})^2 + \frac{\pi^2 J}{ 4 N^2 } \right ]\:\:\:\: .
\label{model6}
\eneq
is the energy part associated to the bare HS Hamiltonian,
and
\beq
q_\ell (m_{\ell}) = \frac{\pi}{2} - \frac{2 \pi}{N} \left[ m_\ell +
\frac{1}{2} \left( K - \ell + \frac{1}{2} \right) \right] \;\;\;\;
\label{model5}
\eneq
denote the spinon (pseudo)~momenta with $0 \leq m_1 \leq \ldots \leq m_K \leq M$.

\begin{figure}[t]
\begin{floatingfigure}[r]{0.5\textwidth}
\scalebox{0.85}{\includegraphics{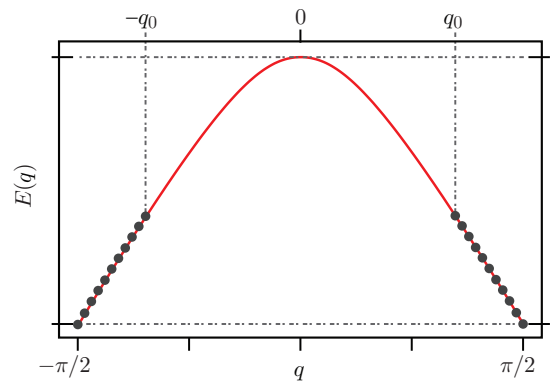}}
\caption{Single-spinon level occupancy corresponding to the state
$ | \Phi_{\rm gnd} , K \rangle$. The spinons symmetrically
 populate low-energy states around the corners of the Brillouin
 zone (which ranges from $- \pi / 2$ to $\pi / 2$). The
 corresponding ``spinon Fermi surface''  is composed by two
 disconnected 'Fermi' points (the largest values of the momenta at which we can create a spin pair), at
 $\pm q_0 = \pm \frac{\pi}{2} \left( 1 - \frac{K}{M} \right)$.}
\label{condenspinon}
\end{floatingfigure}
\end{figure}

\noindent
According to Eqs.~(\ref{model7})-(\ref{model5}), the
minimum energy $K$-spinon fully polarized state is realized for
$m_1 = \ldots = m_{K/2}=M$ and
$m_{K/2 + 1} = \ldots = m_K = 0$, which corresponds to the condensing spinons
being equally distributed at the ends of the single-spinon
Brillouin zone, as depicted in Fig.\ref{condenspinon}.
When all the spins in $ | \Phi_{\rm gnd} , K \rangle$ are
polarized, that is, when $K=N$, the ground state of the system is
given by the simple factorized state  $ | \Phi_{\rm gnd} , N
\rangle = | \downarrow \rangle^{\otimes N}$.
In the language of spinons,
$ | \Phi_{\rm gnd} , N \rangle$ corresponds to a state with the
single-spinon Brillouin zone completely filled.
In fact,  the
``saturated state'' $ | \Phi_{\rm gnd} , N \rangle$  becomes the
ground state of the system as soon as $h>h_s$, where  $h_s$ is the
``saturation field''~(to be estimated below).
The interval $0 \leq
h \leq h_s$ is, therefore, divided into $N/2$ regions labelled by the number
of condensing spinons.
The $n^{\rm th}$ region, with $2n$ spinons~($n=0, \ldots , N/2 - 1$), ends with the energy crossing
between $E_{\rm gnd} [ N , K =2n, h ]$ and $ E_{\rm gnd} [ N , K=2n+2, h]$.
The width of each interval becomes smaller, and the
set of crossing points becomes denser, as the number of spins
increases, which gives rise to a quantum instability similar to
those found for the Dicke and the $XX$~models~\cite{plastina}.

\begin{figure}[t]
\scalebox{0.85}{\includegraphics{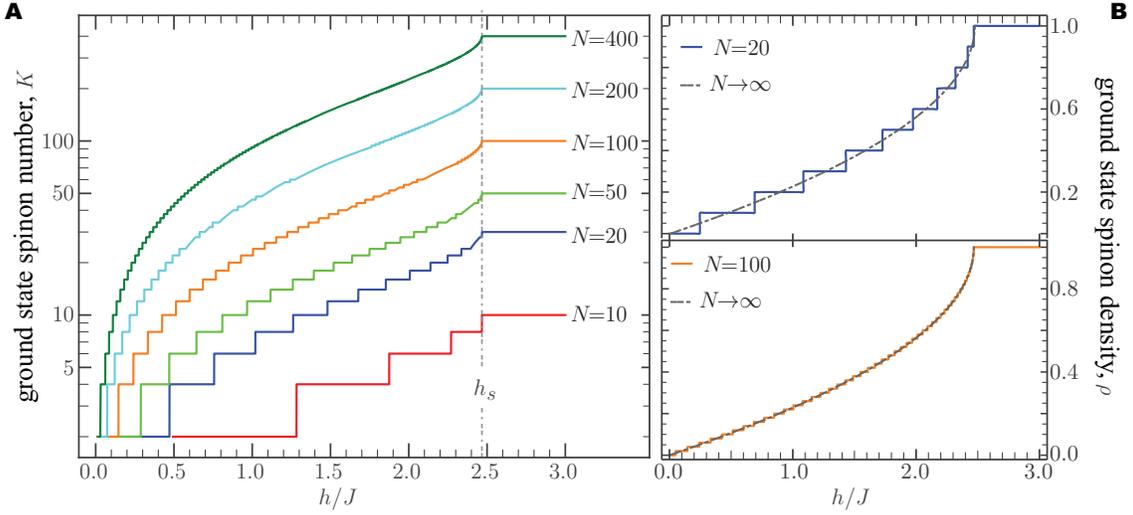}} \caption{\textbf{A}:
Number of condensed spinons, $K [ N , h]$, {\it vs.} $h/J$, for
different values of $N$ ranging from $10$ to $400$. \textbf{B}:
Density of condensed spinons $\rho [ N , h]$, {\it vs.} $h/J$, for
$N=20$ and $N=100$; the thermodynamic expression~(\ref{model8}),
holding exactly for $N \to \infty$, is also shown for comparison.}
\label{kvsH}
\end{figure}

Let $K [ N , h ]$ be the number of condensing spinons, obtained with the minimization procedure
outlined above. The variation of  $K [ N , h ]$
with respect to $h/J$ for different values of $N$ is
reported in Fig.~\ref{kvsH}\textbf{A}, where it is shown
that, for large enough $N$,
the steps in $K [ N , h ]$ are smoothed down.
This implies that the spinon density $K [ N , h ] / N$,  shown in Fig.~\ref{kvsH}\textbf{B},
can be approximated with a
continuum function $\rho_K ( N , h)$, whose dependence on $h$ at
fixed $N$ is given by
\beq \rho_K ( N , h ) = 1 - \sqrt{ 1 - \frac{4h}{\pi^2 J} +
\frac{4}{3 N^2}} \:\:\:\:.
\label{model8}
\eneq
\noindent It is worthwhile noticing that $\rho_K ( N , h )$
is independent of $N$, as $N \to \infty$, which is also displayed in Fig.~\ref{kvsH}\textbf{B}.
The value of the saturation field $h_s$ can be derived from Eq.~(\ref{model8}), by
requiring that $\rho_K ( N , h ) = 1$, for $h > h_s$.
As a result,
one obtains $h_s \approx \pi^2 J \left( \frac{1}{4} + \frac{1}{3
N^2} \right)$, which, for large enough $N$~($\geq 10$), gives $h_s
\approx  \pi^2 J / 4$. Accordingly, the number
of condensed spinons as a function of the external field is given
by
\beq K [ N , h ] = \biggl\{ \begin{array}{l}
N - \frac{2N}{\pi} \sqrt{\frac{h_s - h}{J}} \:\: {\rm for} \: h \leq h_s \\
N \:\:\: {\rm for} \: h > h_s
\end{array}
\:\:\:\ . \label{model9} \eneq
\noindent This can be regarded as a ``thermodynamic limit formula'', which is
confirmed by the plots of Fig.~\ref{kvsH}\textbf{B}.

\section{Ground state entanglement}
\label{entangle}

This section is concerned with the evaluation of two kinds of
entanglement measures, both depending on the external magnetic
field and on the number of spins. One of these is the one-tangle
$\tau_1(N,h)$, which quantifies the amount of entanglement shared
by each spin with the rest of the system; the other is the
concurrence of any pair of spins $C(N,h,r)$ that also depends on
the relative distance between the spin sites
$r=\alpha-\alpha^{\prime}$~($\alpha \neq
\alpha^{\prime}=0,\ldots,N-1$), because of the translational
invariance of the system (see also \cite{CKW}).

Using the concurrence, the two-tangle $\tau_2(N,h)  = \sum_r C^2
(N,h,r)$ and then the entanglement ratio $R(N,h)=\tau_2(N,h) /
\tau_1(N,h)$ will be evaluated. They will provide information
about the fraction of entanglement shared by pairs with respect to
the total~(i.e. both bipartite and multipartite) entanglement to
which each spin participates. In order to evaluate these
entanglement measures, the mean values of the spin operators and
of the two-spin correlation functions are needed on the state $ |
\Phi_{\rm gnd} , K  \rangle $.

The technical steps required to obtain these quantities are
reported in the appendix~\ref{spinprop}, where their explicit
expressions are given, and where the thermodynamic limit is
thoroughly discussed. In particular, all of the one- and two-spin
properties  can be determined analytically for any value of the
external field $h$, and for any given number of spins $N$, except
for the longitudinal correlation function. The latter is obtained
numerically at specific values of $h$~(see appendix~\ref{spinprop}
for the details).

\subsection{One-tangle}
\begin{figure}[tbp]
\vskip -6pt
\begin{floatingfigure}[r]{0.45\textwidth}
\scalebox{0.8}{\includegraphics{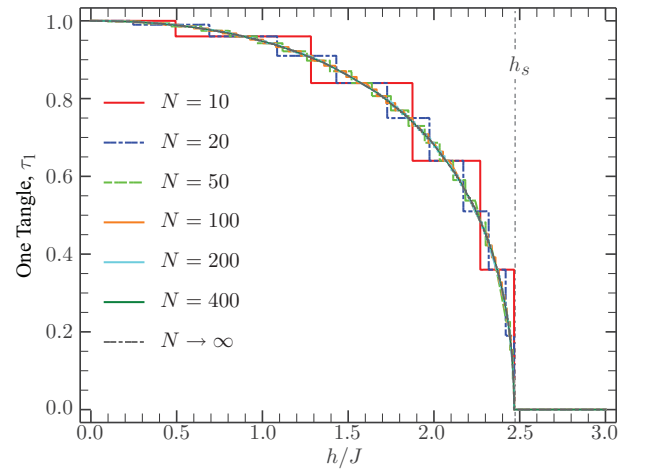}}
\caption{One-tangle $\tau_1$ {\it vs.} $h / J$ for several values
of $N$. For $N > 100$, the plot is well fitted by the
thermodynamic limit formula~(\ref{ent3}).} \label{onetangle}
\end{floatingfigure}
\vskip -22pt
\end{figure}
To compute the one-tangle, one needs to obtain the average value
of the three cartesian components of each spin $S^j_\alpha$, with $j=x,y,z$, on the ground state~\cite{amicorep}.
\noindent Since the applied
magnetic field does not break the rotational symmetry around the
$z$-axis, one gets
\beqs
\beq
\langle  \Phi_{\rm gnd} , K  | S^x_\alpha | \Phi_{\rm gnd} , K  \rangle  =
\langle  \Phi_{\rm gnd} , K  | S^y_\alpha | \Phi_{\rm gnd} , K  \rangle  = 0
\;\;\; ,
\eneq
and
\beq
\langle  \Phi_{\rm gnd} , K  | S^z_\alpha | \Phi_{\rm gnd} , K  \rangle =
-\frac{ K [ N , h ]}{2 N}  \equiv - \bar{m}
\;\;\;,
\label{ent1}
\eneq
where $-\bar{m}$ coincides with the average magnetization of the ground state.
\eeqs
\noindent
Consequently, the one-tangle has the form
\beq \tau_1(N,h) = 1 - 4 \vert \langle  \Phi_{\rm gnd} , K  | S^z_\alpha |
\Phi_{\rm gnd} , K  \rangle \vert^2 = 1 -\frac{K [ N , h ]^2}{N^2}
\:\:\:\: . \label{ent2} \eneq
\noindent
In Fig.~\ref{onetangle}, $\tau_1$ is shown as a function of $h /
J$, for different values of $N$. Similarly to the plots shown in Fig.~\ref{kvsH}\textbf{B}, it is
worthwhile noticing that, for $N > 100$, the
curves are well approximated by the analytical formula
\beq
\tau_1 (N \to \infty,h) = \biggl\{ \begin{array}{l} \frac{4}{\pi} \sqrt{
\frac{h_s - h}{J}} - \frac{4}{\pi^2} \left( \frac{h_s - h}{J}
                         \right) \;\;\; , \;\; {\rm for} \: h \leq h_s \\
0 \;\;\; , \;\; {\rm for } \: h > h_s
                        \end{array}
\:\:\:\: ,\label{ent3}
\eneq
\noindent obtained in
the thermodynamic limit, and yielding the dominant
behavior \beq\tau_1 (N \to \infty, h\rightarrow h_{s}^-) \sim \left(
h_{s}-h\right) ^{1/2}\:\:\:\: .\label{ent3B}
\eneq

\subsection{Bipartite entanglement}

By taking into account the symmetries of the Hamiltonian, it is
easy to show that the two spin reduced density matrix has the so
called $X$-form and that the concurrence $C (N, h , r )$ can be
expressed in terms of the two-point correlation functions
\beq
g^{jj^{\prime}} (N, h, r) =
\langle  \Phi_{\rm gnd} , K | S_{\alpha = 0}^j S_{\alpha^{\prime} = r}^{j^{\prime}} |   \Phi_{\rm gnd} , K
\rangle\;\;\;,\label{corrfunc}
\eneq in which $j,j^{\prime}=x,y,z$~\cite{amico}.
In particular, when two spins are at a distance $r$
from each other, one gets $C (N, h, r ) = {\rm max} \{ 0 , C_a (N, h, r) \}$,
where
\beq C_a (N, h, r) = 4 \sqrt{ [g^{xx} (N, h, r) ]^2 + [g^{xy} (N, h, r) ]^2 } - 2 \sqrt{ [ 1 / 4 + g^{zz} (N, h, r) ]^2 - \bar{m}^2 }
\:\:\:\: \label{ent4}
\eneq
denotes the antiparallel concurrence.
\noindent Since the latter depends on the longitudinal spin-spin correlation function,  $g^{zz}(N, h, r)$, its expression
can be obtained analytically only in the range of $h$ for which such a function is
is known~(see
appendix~\ref{spinprop}), and is otherwise computed numerically.

For $h=0$, using $ g^{jj} (N,h=0,r) \equiv g_0 (r) $, for each
$j=x,y,z$, and $g^{xy} (N,0,r) = 0$, one obtains the zero-field
concurrence: \beq C (N,0,r )  = {\rm max} \{ 0 , 4 | g_0 ( r ) | -
2 \sqrt{ 1 / 4 + [ g_0 ( r ) ]^2 } \} \:\:\:\: . \label{ent5}
\eneq An explicit evaluation shows that, for any $N$, $C_a (N,h=0
, r)$ is always negative, except for the cases $r = 1$ and
$r=N-1$. The same property holds in the thermodynamic limit where
$g_0 ( r )$ takes the expression:~\cite{haldaneshastry} \beq g_0 (
r ) \approx \frac{ (-1)^r }{16 \pi} \: \int_{ - \pi/2 }^{\pi/2 }
\: d q_\alpha \: d q_\beta \: \frac{ | q_\alpha - q_\beta | e^{ -
i ( q_\alpha + q_\beta ) r } }{ \sqrt{\pi^2/4
 - q_\alpha^2} \: \sqrt{ \pi^2/4 - q_\beta^2}  }
\:\:\:\:, \label{spcor6a} \eneq \noindent and the condition $C_a
(N,h=0,r) > 0$ is satisfied only for $r=1$. Therefore, for $h =
0$, only the concurrence between spins lying onto nearest
neighboring sites is different from zero, as it could be expected
for an isotropic spin liquid state.

\begin{figure}
\scalebox{0.85}{\includegraphics{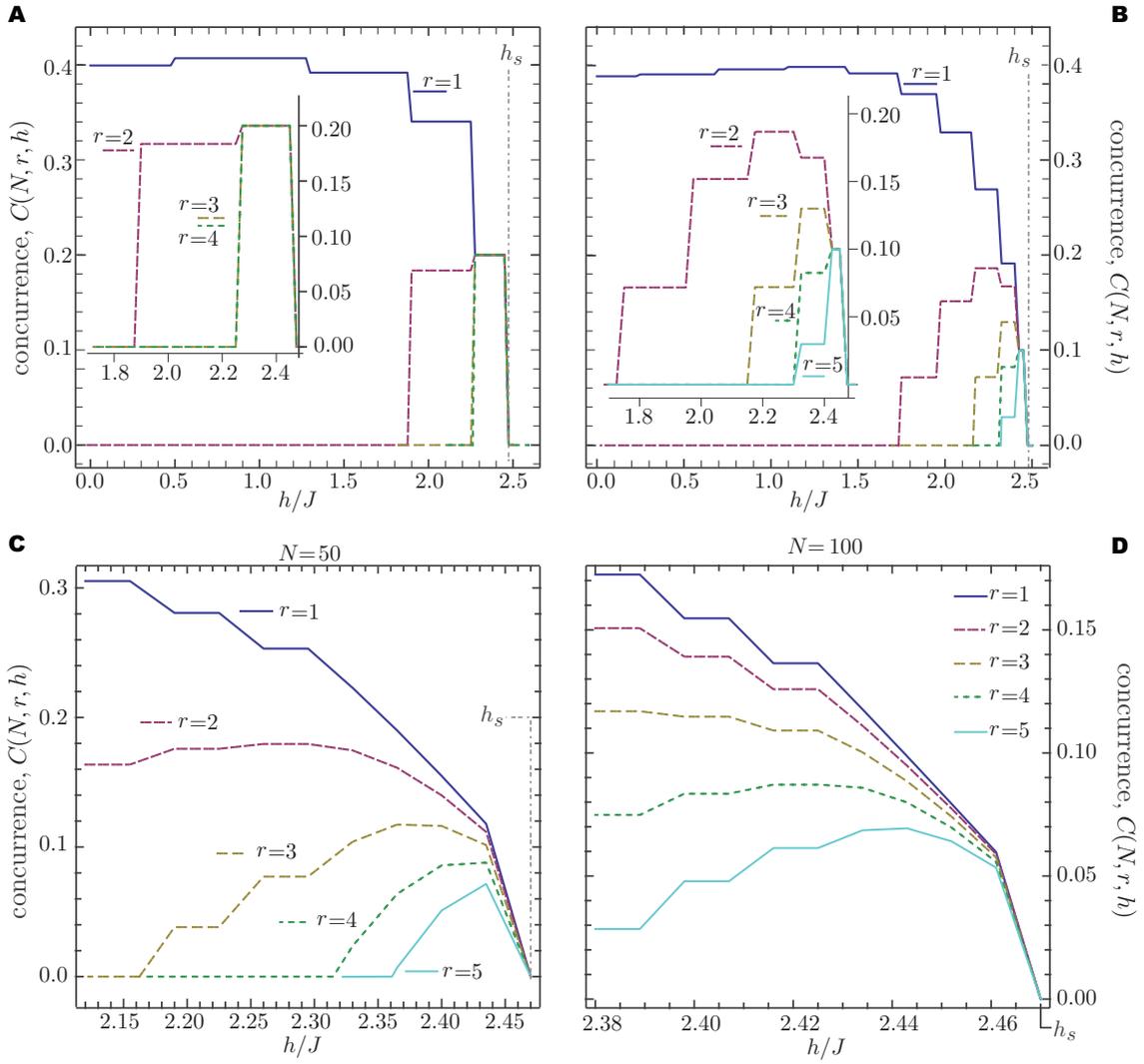}} \caption{$C(N,h,r)$
{\it vs.} $h$ at fixed $r$ and for various values of $N$. Upon
increasing $h$, $C (N,h,r=2)$ can be made $>0$, thus yielding
nonzero second neighbor concurrence. At a higher value of $h$,
$C(N,h,r \geq 3)$ becomes $> 0$, as well.} \label{fhconcnum}
\end{figure}

Due to the lack of an explicit analytical formula for $g^{zz}
(N,h,r)$, the concurrence in the intermediate range $0 < h < h_s$
needs to be computed numerically. A result in the whole range can
be easily found in the case $N \leq 20$; in particular,
Figs.~\ref{fhconcnum}\textbf{A} and~\ref{fhconcnum}\textbf{B} show
the behavior of the concurrence {\it vs.} $h$, at fixed $r$, for
$N=10$ and $N=20$, respectively. One clearly sees that the
switching on of the magnetic field leads to an enlargement of the
entanglement range, as $C (N,h,r)$ becomes positive also for
$r>1$.

\begin{figure}
\scalebox{0.75}{\includegraphics{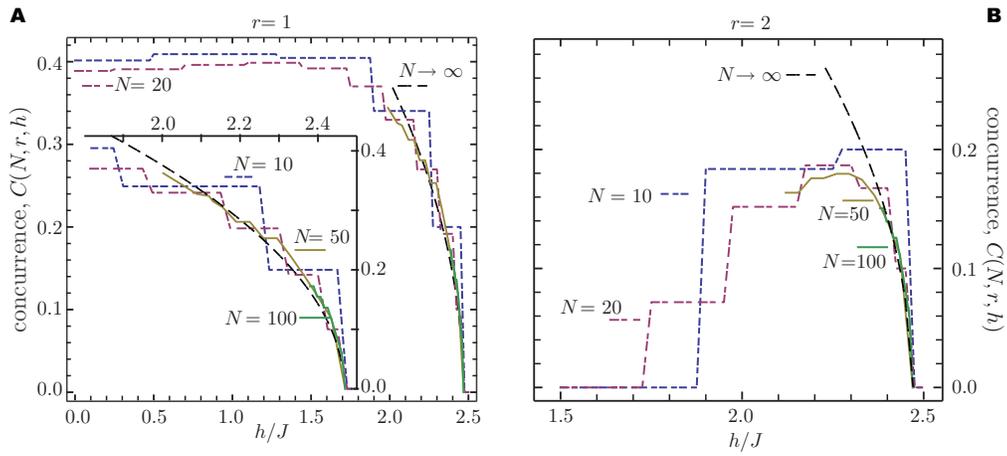}} \caption{Behavior of
$C(N, r,h )$ for $h \to h_s$ with  $r=1$ (left panel) and $r=2$
(right panel) and for $N=10,20,50,100$, and $N \to \infty$
(thermodynamic limit). Upon increasing $h$, the entanglement range
increases while at $h_s$ all the curves show the universal
behavior of Eq.~(\ref{ent6})} \label{highh}
\end{figure}

For $N \geq 50$, it is still possible to determine $C (N,h,r)$
exactly but only at values of $h$ close to the saturation field.
This is shown in plots of Figs.~\ref{fhconcnum}\textbf{C}
and~\ref{fhconcnum}\textbf{D}, where the behavior of the
concurrence near $h_s$ is shown for $N=50,100$ and $r=1-5$. In
fact, when $h \sim h_s$, the leading contribution to $C(N,r,h )$
can be calculated by thermodynamic arguments, putting together the
relationships~(\ref{spcor6}) and~(\ref{spcor11}) derived in the
appendix~\ref{spinprop}. For relevant values of $r$~($r \neq
0,N$), one obtains \beq C_a (N \to \infty, h, r) = 4 q_0 \left\{ |
\gamma ( q_0 r )| - \frac{1}{\pi} \sqrt{ \left( 1 - \frac{2
q_0}{\pi} \right) {\cal F} [ q_0 , r ] + \left( \frac{2 q_0}{\pi}
\right)^2 {\cal F}^2 [ q_0 , r ] } \right\} \:\:\:\: ,
\label{ent5.1} \eneq \noindent where the functions $\gamma$ and
${\cal F}$ are also defined in the appendix~\ref{spinprop} and
$q_0$, the magnitude of the `Fermi momentum' indicated in
Fig.~\ref{condenspinon}, can be expressed as $q_0 = \frac{\pi}{2}
( 1 - 2 \bar{m})$. As a result, one finds \beq C_a (N \to \infty,
h \sim h_s, r ) \sim \sqrt{ ( h_s - h ) / J} \: \Phi ( q_ 0 , r
)\:\:\:\: , \eneq where $\Phi ( q_0 ,r )$ is a regular function of
$q_0 , r$. In particular, at $h \to h_s$, one gets $q_0 = 0$
implying $\gamma ( 0 ) = \pi/2$ and ${\cal F} [ q_0 \to 0 , r ] =
0$, as it will emerge from the discussion in the
appendix~\ref{spinprop}. Thus, \beq C_a (N \to \infty, h \to h_s,
r ) = \frac{1}{\pi} \: \sqrt{ \frac{h_s - h}{J} } + {\cal O}
\left(  \frac{h_s - h}{J} \right) \;\;\;\; , \label{ent6} \eneq
\noindent independently of $r$. The remarkable collapse of the
curves onto each other for $h \to h_s$, reported in
Fig.\ref{highh} for $r = 1,2$ and for $N = 10,20,50,100$, confirms
the prediction of Eqs.~(\ref{ent5.1}),~(\ref{ent6}).

\begin{figure}[!!h]
\scalebox{0.85}{\includegraphics{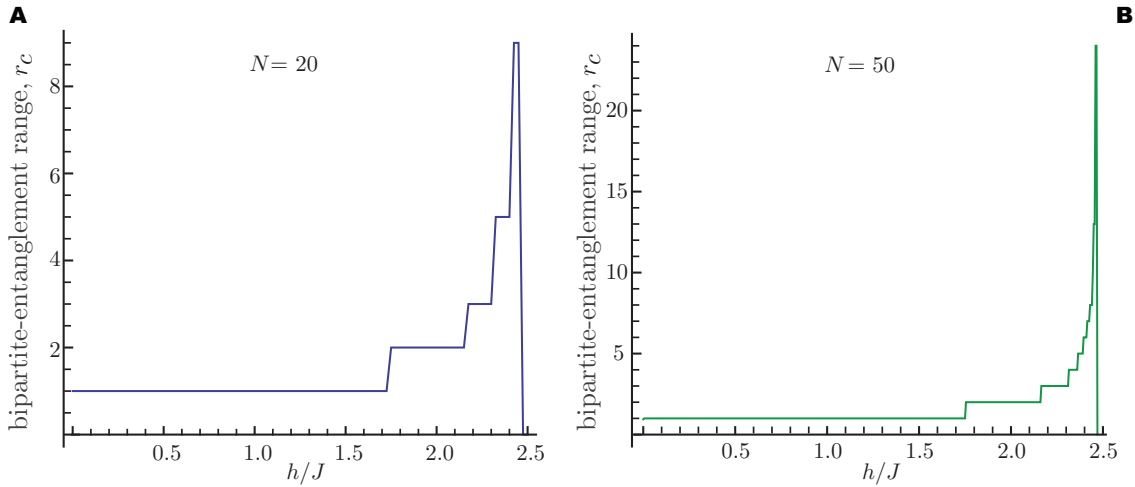}} \caption{Behavior of
the range of the bipartite entanglement \textit{vs} $h/J$; as $h
\to h_s$, all the spinons are entangled and $r_s \to N/2$}
\label{range}
\end{figure}

In general, one can say that by increasing the distance $r$, the
concurrence $C(N,h,r)$ is different from zero only in a smaller and
smaller region of the magnetic field and that the range of
the bipartite entanglement, $r_c$, extends to the whole chain near the
saturation point $h_s$. The behavior of $r_c$ vs $h/J$ is shown in Fig.~\ref{range}.
This is clearly reminiscent of
what happens at factorization points~\cite{luigipaola}.

In accordance with this observation, we show below that near the
saturation field, the bi-partite entanglement is dominant  with
respect to multi-partite ones, as it is the case for factorization
points.

\subsection{Two-tangle and entanglement ratio}
In order to understand wether the entanglement stored in the
ground state has a multi-partite rather than a bi-partite nature,
we consider in this section the ratio between the two-tangle and
the one-tangle defined above. Due to the monogamy relation for
entanglement, this ratio is bound between zero (for purely
multipartite correlations) and one (for bipartite entanglement)
\cite{CKW}.

\begin{figure}[h]
\begin{centering}
\scalebox{0.85}{\includegraphics{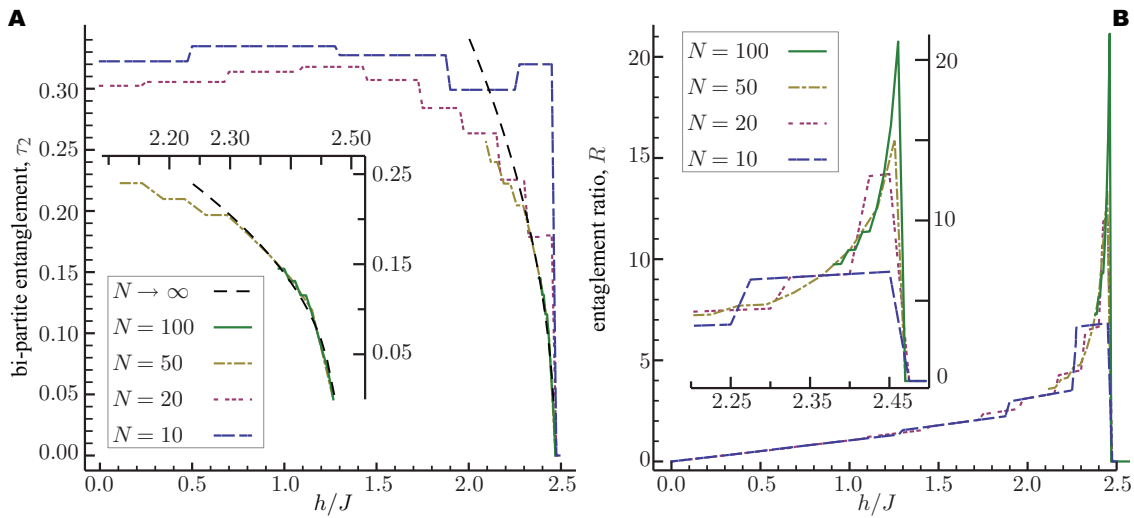}}
\end{centering}
\caption{\textbf{A}: Two-tangle, $\tau_2$ as a function of the
magnetic field. The black dashed line gives the limiting behavior
reported in Eq.(\ref{limitre}). \textbf{B}: Entanglement ratio as
a function of the external field. The plot suggests that near the
saturation point, $\tau_2$ dominates over $\tau_1$. }
\label{ig9}\end{figure}

We first focus on the two-tangle, that gives an overall measure of
the bi-partite entanglement to which a given spin participate. It
is defined by
\begin{equation}
\tau_2(N,h)= 2\sum_{r=1}^{N/2}C(N,h,r) ^{2}\text{,}
\end{equation}%
and its behavior with
respect to the magnetic field is reported in Fig.~\ref{ig9}\textbf{A}.

\noindent When $h \to h_{s}$, the limiting trend~(black dashed line) is
given by
\begin{equation}
\tau_2 (N,h\rightarrow h_{s}) \sim \left( h_{s}-h\right)^{\alpha
}, \label{limitre}
\end{equation}
with an exponent that we estimated, by fitting, to be $\alpha =0.50 \pm 0.02$.
This already implies that, near $h_s$, $\tau_2$ goes to zero as
the one-tangle: $\tau_1 \sim (h_s-h)^{0.5}$.

To better show this fact, we finally consider the entanglement
ratio $R(N,h)=\tau_2(N,h)/\tau_1(N,h)$, whose behavior is reported in
Fig.~\ref{ig9}\textbf{B}. Near the saturation field, the limiting
trend $R(N,h \to h_s) \sim \left( h_{s}-h\right) ^{\beta }$ has an exponent
$\beta$ taking values smaller than $0.03$, which is consistent
with the other estimates given above. More importantly, the plot
shows that at $h \to 0$ the one-tangle dominates, whereas in the
neighborhood of $h_s$  the entanglement is essentially bi-partite.

\section{Concluding Remarks}
\label{close} Typical spin models considered so far to discuss the
behavior of entanglement lie into two main categories: those with
nearest neighbor interaction, such as the Ising model in a
transverse field, and those with long range interaction with spins
residing on a fully connected graph, such as the LMG model
\cite{lmgnuovi} or the uniaxial model \cite{vidal}. In the former
case, \cite{amicorep}, bipartite entanglement decays quickly with
the distance, and only nearest and next to nearest neighbor
entanglement is found to survive. Notable exception are
factorization points around which the concurrence is found to have
an infinite range \cite{luigipaola}. In collective spin models, on
the other hand, the qubit-qubit entanglement is strictly zero in
the thermodynamic limit, and its presence in the ground state is
only due to finite size effects. In particular, the concurrence is
found to scale as $(N-1)^{-1}$, where $N$ is the number of spins.

We have found that the inverse square (statistical) interaction
among spins produces a quite different behavior of the
entanglement as a function of the distance. In particular, we have
shown that, in the absence of external magnetic field, $h=0$, the
system only displays nearest neighbor concurrence, while the
entanglement has essentially a multi-partite nature; while, by
increasing $h$, the entanglement becomes essentially bi-partite
and the range of the concurrence increases. For a large enough
$h$, the spin system eventually saturates, entering a fully
polarized phase described by a completely separable ground state.
We have discussed in particular the behavior of the entanglement
near this saturation point, both for a finite size system and in
the thermodynamic limit. In particular, we found that the
divergence of the range of the bipartite entanglement  is governed
by the same exponent ($|h-h_s|^{1/2}$) of the ordinary isotropic
Heisenberg model, consistently with the fact that the latter
belongs to the same universality class as the Haldane-Shastry
model.

Since the inverse square interaction among spins in the HS model
is essentially of statistical nature, the fact that the
entanglement range tends to diverge near the saturation can be
compared with the behavior of entanglement for other systems of
indistinguishable free particles. As mentioned in the introductory
section, the spin entanglement of a free electron gas is different
from zero within a distance of the order of the Fermi wavelength.
On the other hand, we have shown that in the presence of a large
number of semions (i.e., near $h=h_s$), the range of the bipartite
entanglement extends to comprise the entire system. This behavior
is ultimately due to the fractional statistics: by increasing the
value of the external field, more and more spinons {\it can be
added to the system} with smaller and smaller momenta, starting
from the edge of the Brillouin zone and going towards its center.
This implies that the only length scale of the system is the
'Fermi' wavelength $q_0^{-1}$, which, because of the semionic
statistics, diverges when the saturation point is approached. In
a sense, this behavior interpolates between the cases of
entanglement among bosons and fermions. Free bosons can be always
in a factorized state because there is no correlation between
their relative distance and their polarization; for free fermions
the range of bipartite entanglement between their spin is
spatially limited to a finite Fermi length. For the semionic gas a
Fermi length  does exist, but it diverges at $h_s$. Therefore,
because of their quantum statistics  an overlap of the spinon wave
functions (whose extension is roughly given by $\hbar/q_0$) can
take place at $h_s$, ultimately making the saturation phenomenon a
result of spinon condensation.

\appendix

\section{Magnetization and Spin correlation functions on $ | \Phi_{\rm gnd} , K \rangle$}
\label{spinprop}

This appendix contains the one- and two-spin correlation functions
on $ | \Psi_{\rm gnd} , K \rangle$. These are the basic bricks to
build the single- and two-spin density matrices needed to evaluate
the entanglement. When possible, the calculations are carried out
analytically on a finite chain and the thermodynamic limit is
systematically worked out by sending $N \to \infty$, while keeping
$K/N$ constant. When exact analytical results are lacking, namely
in the calculation of the longitudinal two spin correlation
functions, the corresponding quantities have been evaluated
numerically.

\subsection{One-spin average values}

Since $ | \Phi_{\rm gnd} , K \rangle$ is an eigenstate of $S^z =
\sum_{ \alpha =1}^N S_\alpha^z$, with total eigenvalue $-K /2$,
one readily obtains $ \langle  \Phi_{\rm gnd} , K | S_\alpha^x |
 \Phi_{\rm gnd} , K \rangle = \langle  \Phi_{\rm gnd} , K | S_\alpha^y |
\Phi_{\rm gnd} , K \rangle = 0$. Moreover, because of
the translational invariance on the lattice, one also has
\beq \langle  \Phi_{\rm gnd} , K | S_\alpha^z |
 \Phi_{\rm gnd} , K \rangle  = - \frac{1}{2} + M   \sum_{ z_2 , \ldots , z_M }
 | \Phi_K ( 1 , z_2 , \ldots , z_M )|^2 =  - \frac{K}{2N}
\:\:\:\: . \label{spcor1} \eneq \noindent
The thermodynamic limit
is defined as $N \to \infty$, with $K / 2 N = \bar{m} = \text{constant}$. Therefore, for $N\to \infty$, one
can write
\beq \langle  \Phi_{\rm gnd} , K | S_\alpha^z |
 \Phi_{\rm gnd} , K \rangle  = - \bar{m}
\:\:\:\: .
\label{spcor2}
\eneq
\noindent

\subsection{Transverse two-spin correlation functions}

The transverse two-spin correlation functions, introduced in Eq.~(\ref{corrfunc}), are here denoted $g^{jj^{\prime}} ( r )$, for $j,j^{\prime} = x,y$ and $r=1,\ldots,N-1$, where for notational simplicity the dependence upon $N$ and $h$ has been omitted.
These functions
can be expressed in terms of the
correlation functions of the operators $S_\alpha^\pm  = S_\alpha^x
\pm i S_\alpha^y$ as
\beq g^{xx} ( r ) = g^{yy} ( r ) = \bar{m} \delta_{r,0} + \Re e [
g_\parallel ( r ) ] \;\;\; , \;\; g^{xy} ( r ) = - g^{yx} ( r ) =
- i \bar{m} \delta_{r, 0} - \Im m [ g_\parallel ( r ) ] \:\:\:\: ,
\label{spcor3} \eneq
\noindent with $ g_\parallel ( r ) = \langle
\Phi_{\rm gnd} , K | S_0^+ S_r^- |
 \Phi_{\rm gnd} , K \rangle $.
The auxiliary function $ g_\parallel ( r ) $
can be exactly calculated by noticing that the act of flipping down
one more spin in the state $ | \Psi_{\rm gnd} , K \rangle $ is
equivalent to the creation of a pair of $\downarrow$-spinons ``on
top of each other''. Following the technique developed in
Ref.~\cite{bgl}, to compute the function  $ g_\parallel ( r )$ for
$K = 0$, one obtains
\beq
g_\parallel ( r ) =  \sum_{ m_\alpha = 0}^{M-1} \sum_{ m_\beta = 0}^{m_\alpha}
e^{ 2 \pi i  (  m_\alpha + m_\beta + 1 + K /2 ) r / N} \:  \chi_{m_\alpha , m_\beta }
\:\:\:\: ,
\label{spcor4}
\eneq
\noindent
for a generic value of $K$, with
\beq
\chi_{m_\alpha , m_\beta} = \frac{ ( m_\alpha - m_\beta + 1 /2 )}{ 2 N } \:
\frac{ \Gamma [ M - m_\alpha - \frac{1}{2} ]}{ \Gamma [ M - m_\alpha ]} \:
\frac{ \Gamma [ m_\alpha + 1] }{ \Gamma [ m_\alpha + 3 / 2 ] }
\: \frac{ \Gamma [ m_\beta + 1 / 2 ]}{ \Gamma [ m_\beta + 1]} \:
\frac{ \Gamma [ M - m_\beta ]}{ \Gamma [ M - m_\beta + 1/2]}
\:\:\:\: .
\label{spcor5}
\eneq
\noindent
To extract the asymptotic form of $g_\parallel ( r )$ in the thermodynamic
limit, the Stirling's formula
$\Gamma [ z ] \approx \sqrt{\pi} ( z-1)^{z-\frac{1}{2}} \: e^{ - ( z-1)}$ is used
to approximate $\chi_{m_\alpha , m_\beta}$ . As
a result, Eq.~(\ref{spcor4}) becomes
\beq
g_\parallel ( r ) \approx \frac{ (-1)^r }{16 \pi} \:
\int_{ - q_0  }^{ q_0 } \:
d q_\alpha \: d q_\beta \: \left[ \frac{ | q_\alpha - q_\beta | e^{ - i ( q_\alpha + q_\beta ) r } }{ \sqrt{q_0^2
 - q_\alpha^2} \: \sqrt{ q_0^2 - q_\beta^2}  } \right]
\:\:\:\:, \label{spcor6}
\eneq
\noindent where $q_0$ is the 'Fermi momentum' introduced in Fig.~\ref{condenspinon}.
An alternative formula for
Eq.~(\ref{spcor6}) is obtained by using the integration variables $\phi_\alpha$ and $\phi_\beta$,
calculated from $q_{\alpha ( \beta )} = q_0
\sin ( \phi_{ \alpha ( \beta )})$. In this way, one obtains
\beq
g_\parallel ( r ) \approx (-1)^r q_0 \: \gamma ( q_0 r )
\:\:\:\: ,
\label{spcor7}
\eneq
\noindent
with
\beq
\gamma ( p ) = \frac{ 1}{16 \pi} \:
\int_{ - \pi / 2  }^{ \pi / 2 } \: d \phi_\alpha \: d \phi_\beta \:
| \sin ( \phi_\alpha ) - \sin ( \phi_\beta ) | \: e^{ i p [ \sin ( \phi_\alpha )
+ \sin ( \phi_\beta )]}
\:\:\:\: .
\label{spcor8}
\eneq
\noindent Eq.~(\ref{spcor7})  is particularly useful for $h \sim h_s$ (that is,
for $q_0 \to 0$).

\subsection{Longitudinal two-spin correlation functions}

The analytical computation of the longitudinal two-spin
correlation functions, also defined in Eq.~(\ref{corrfunc}) for $j=j^{\prime} = z$ and $r=1,\ldots,N-1$, is
quite a formidable task not yet fully addressed~\cite{haldanemagnetic,kuramotomagnetic,katsura}.
Even though in this paper $g^{zz} ( r )$ is computed numerically for
finite $N , K$, its  leading contributions in $q_0$, as well as
its exact analytical expression for $h \to h_s$, are
analytically determined.
Indeed, by using the
explicit form of $|   \Phi_{\rm gnd} , K \rangle$ given by
Eq.(\ref{model2}), one obtains
\beq
g^{zz} ( r ) = \frac{1}{4} - \frac{q_0}{\pi} ( 1 - \delta_{r , 0 } ) + M ( M -1 )
\sum_{ \{ z_3 , \ldots , z_M \} \in S^N} \: | \Phi_K ( 1 , e^{ 2 \pi i r / N } , z_3 , \ldots , z_M ) |^2
\:\:\:\: ,
\label{spcor9}
\eneq
\noindent
where $S^N$ is set of the $N^{\rm th}$ roots of the unity. Now, as pointed out in Ref~\cite{bgl}, one can write
\beq
 \sum_{ \{ z_3 , \ldots , z_M \} \in S^L} \: | \Phi_K ( 1 , e^{ 2 \pi i r / N } , z_3 , \ldots , z_M ) |^2
= L^{M-2} \: \prod_{ j = 3}^M \: \oint_\gamma \frac{d z_j}{2 \pi i
z_j } \: | \Phi_K ( 1 , e^{ 2 \pi i r / N } , z_3 , \ldots , z_M )
|^2 \:\:\:\: , \label{spcor10}
\eneq \noindent for any
integer $L$, in  which $\gamma$ is
the unit circle in the complex plane.
By
Eq.(\ref{spcor10}), one readily gets
\beq
g^{zz} ( r ) = \frac{1}{4}  - \frac{q_0}{\pi} (1- \delta_{r,0} ) + \left(\frac{2 q_0}{ \pi} \right)^2 \:
{\cal F} [ q_0 , r ]
\:\:\:\: .
\label{spcor11}
\eneq
\noindent
In this relationship, the function ${\cal F} [ q_0 ,   r ]$ has the definition
\beq
{\cal F} [ q_0 ,   r ] = \frac{2^{M} M ( M - 1)}{(2M )! (2M)^M}
| 1 - e^{ 2 \pi i r / N}  |^4
\sum_{ \{ z_3 , \ldots , z_M \} \in S^{2M}} \:  \prod_{i < j =3}^M \: | z_i - z_j |^4  \prod_{ j = 3}^M |
1 - z_j |^4 | e^{ 2 \pi i r / N} - z_j |^4
\:\:\:\: ,
\label{spcor12}
\eneq
\noindent
for $M \geq 2$ ($q_0 > 0$), and ${\cal F} [ q_0 ,   r ] = 0$ for $M = 0,1$.
It is straightforward to check that $| {\cal F} [ q_0 ,   r ] | \leq 1$ for any choice of
$q_0$ and $r$. Clearly, this gives a strong constraint on the behavior of
$g^{zz} ( r )$ for $q_0 \to 0$, that is,
\beq
g^{zz} ( r ) = \frac{1}{4}  - \frac{q_0}{\pi} (1- \delta_{r,0} ) + {\cal O} ( q_0^2)\:\:\:\: .
\label{spcor12B}
\eneq
Moreover, for small $M$, the function ${\cal F} [ q_0 ,   r ]$ can be explicitly
evaluated. For instance, for $M = 2$ ($q_0 = 2 \pi / N$), one gets
\beq
g^{zz} ( r ) = \frac{1}{4}  - \frac{q_0}{\pi} (1- \delta_{r,0} ) + \frac{ q_0^2}{ 2 \pi^2}  \: \sin^4
\left( \frac{\pi r}{N} \right)
\:\:\:\: ,
\label{spcor13}
\eneq
\noindent
while, for $M=1$ ($q_0 =  \pi / N$), the result is simply
$g^{zz} ( r ) = \frac{1}{4}  - \frac{q_0}{\pi} (1- \delta_{r,0} )$.

In the absence of an external field, $g^{zz} ( r )$ is exactly
computed by simply using the fact that  the ground
state, $| \Phi_{\rm gnd} , K = 0 \rangle$, is a spin-$0$ spin
singlet \cite{haldaneshastry,bgl}. Accordingly, one gets $g^{zz} ( r ) =
g^{xx} ( r ) = g^{yy} ( r )$, and $ g^{xy} ( r ) = g^{yx} ( r ) =
0$.


\begin{thebibliography}{99}

\bibitem{amicorep} L. Amico, R. Fazio, A. Osterloh, and V. Vedral,
Rev. Mod. Phys. {\bf 80}, 517 (2008); L. Amico and R. Fazio, J. Phys. A: Math. Theor. {\bf 42} 504001 (2009).

\bibitem{Ghirardi} G.C. Ghirardi and L. Marinatto, Optics and Spectroscopy 99, 386 (2005).

\bibitem{Vlatko-fermions} V. Vedral, Central Eur.J.Phys. {\bf 1} 289 (2003).

\bibitem{Zanardi} P. Zanardi, Phys. Rev. A {\bf 65}, 042101 (2002).

\bibitem{Cavalcanti}
K. Eckert, J. Schliemann, D. Bruss, and M. Lewenstein, Annals of Physics {\bf 299}, 88 (2002);
D. Cavalcanti {\it et al.}, Phys. Rev. B {\bf 76}, 113304 (2007).



\bibitem{Les-Houches1999} A. Comtet, {\it et al.}, 'Topological aspects of low dimensional
systems', Les Houches Session LXIX (Springer, Berlin 1998).

\bibitem{Haldane-exclusion}  F. D. M. Haldane, Phys. Rev. Lett. {\bf 66},
        1529 (1991).

\bibitem{Calogero-statistics} M. V. Murthy and R. Shankar, Phys. Rev. B {\bf 60}, 6517 (1999).

\bibitem{partitions}The concepts on entanglement entropy between particle partitions in itinerant systems are reviewed in 
M. Haque, O. S. Zozulya, and  K Schoutens,  J. Phys. A: Math. Theor. {\bf 42} 504012 (2009).
\bibitem{haldaneshastry} F. D. M. Haldane, Phys. Rev. Lett. {\bf 60},
        635 (1988); B. S. Shastry, Phys. Rev. Lett. {\bf 60},
        639 (1988).

\bibitem{Haldane-ideal} F. D. M. Haldane, {\it Proceedings of the 16th. Taniguchi
Symposium on Condensed Matter Physics, Kashikojima, Japan}, A.
Okiji and N. Kawakami eds. (Springer, Berlin 1994).

\bibitem{toadd1} A. M. Polychronakos, Phys. Rev. Lett. {\bf 70}, 2329 (1993).

\bibitem{haldanemagnetic} J. C. Talstra and F. D. M. Haldane, Phys. Rev. {\bf B 50}, 6889
                          (1994); Phys. Rev. {\bf B 54}, 12594 (1996).


\bibitem{faddeev} J.des Cloizeaux and J. J. Pearson, Phys. Rev. {\bf 128}, 2131 (1962);
                  L. D. Fadeev and L. A. Takhtajan, Russian Math.
                  Surveys {\bf 34} 11 (1979); Phys. Lett.  {\bf 85A}, 375 (1981).


\bibitem{bgl} B. A. Bernevig, D. Giuliano, and R. B. Laughlin, Phys. Rev. Lett. 86, 3392 (2001); Phys. Rev.
              {\bf B 64}, 024425 (2001).



\bibitem{plastina} V. Buek, M. Orszag, and M. Roko, Phys. Rev. Lett. 94, 163601 (2005);
                   W. Son, L. Amico, F. Plastina, and V. Vedral, Phys. Rev. {\bf A 79}, 022302 (2009).

\bibitem{CKW}  V. Coffman, J. Kundu, W. K. Wootters,
    Phys.Rev. A {\bf 61}, 052306 (2000); T. J. Osborne and F. Verstraete, Phys. Rev. Lett. {\bf 96}, 220503
    (2006).


\bibitem{amico}  L. Amico, A. Osterloh, F. Plastina, R. Fazio, and G. M. Palma, Phys. Rev. {\bf A 69},
                 022304 (2004).


\bibitem{luigipaola} L. Amico,  F. Baroni, A. Fubini, D. Patan\`{e}, V. Tognetti, and
P. Verrucchi, Phys. Rev. A {\bf 74}, 022322 (2006).


\bibitem{lmgnuovi} P. Ribeiro, J. Vidal, and R. Mosseri,
Phys. Rev. Lett. {\bf 99}, 050402 (2007); P. Ribeiro, J. Vidal,
and R. Mosseri, Phys. Rev. E {\bf 78}, 021106 (2008).


\bibitem{vidal} J. Vidal, Phys. Rev. A \textbf{73}, 062318 (2006).

\bibitem{kuramotomagnetic} M. Arikawa, Y. Saiga, T. Yamamoto, and Y. Kuramoto,
                           Physica B 281, 823 (2000).

\bibitem{katsura} H. Katsura and Y. Hatsuda, J. Phys. {\bf A 40},13931 (2007).


\end{thebibliography}
\end{document}